\documentclass[final,5p,times,twocolumn]{elsarticle}

\usepackage{amssymb}
\usepackage{amsmath}

\usepackage{amsthm}
\usepackage[colorlinks,citecolor=blue,linkcolor=blue]{hyperref}
\newtheorem{definition}{Definition}

\newtheorem{remark}{Remark}
\newtheorem{theorem}{Theorem}
\newtheorem{lemma}{Lemma}
\newcommand{\letsymb}[1]{#1}
\usepackage{dsfont}
\usepackage{xcolor}
\newcommand{\e}{\mathrm{e}}
\newcommand*{\Conflictofinteres}[0]{\textbf{\\ Conflict of interest}\textendash}
\newcommand*{\DataAvailability}[0]{\textbf{\\Data Availability}\textendash}
\newcommand*{\Acknowledgement}[0]{\textbf{\\Acknowledgements}\textendash}
\usepackage{physics}

\journal{Annals of Physics}
\usepackage{lineno} 
\begin{document}
	
	\begin{frontmatter}
		
		\title{Unification of  stochastic matrices and quantum operations for N-level systems}
	
		\author{Bilal Canturk} 
		\ead{bcanturk@mersin.edu.tr} 
		\affiliation{organization={Department of Physics, Mersin University},
			city={Mersin},
			postcode={33343}, 
			country={Turkiye}}
		
		\begin{abstract}
			The time evolution of the one-point probability vector of stochastic processes and quantum processes for $N$-level systems have been unified. Hence,  quantum states and quantum operations can be regarded as generalizations of the one-point probability vectors and stochastic matrices, respectively. More essentially, based on the unification, it has been proven that completely positive divisibility (CP-divisibility) for quantum operations is the natural extension of the Chapman-Kolmogorov equation. It is thus shown that CP-divisibility is a necessary but insufficient condition for a quantum process to be specified as Markovian. The main results have been illustrated through  a dichotomic Markov process.
		\end{abstract}
		
	\end{frontmatter}
	\section{Introduction}
	The conceptual and theoretical differences between classical theories and quantum theory show its effect also in other branches of physics such as stochastic and quantum processes \cite{Rivas2014a,Weissman-Breuer-Vacchini2015, Vacchini-Smirne-Breuer2011}. For instance, for the time evolution within finite-dimensional space,  while the main elements of stochastic processes are probability vectors and stochastic matrices \cite{Feller1970V2, Gardiner2009}, those of quantum processes are quantum states and quantum operations \cite{Nielsen-Chuang2010, Watrous2018}. These fundamental differences have thus caused different reformulations of some basic elements of stochastic processes in quantum processes such as the definition of Markovianity. Indeed, there are various definitions of quantum Markovianity  that differ from each other and give rise to inconsistent conclusions about a particular system (for a review of the definitions, see e.g., \cite{Rivas2014a, Breuer2016, Li2018a} and the references therein). However, none of them construes a satisfactory connection with the classical definition of Markovianity \cite{Li2018a}. To develop such a connection, it is an essential requirement to extend consistently the Chapman-Kolmogorov equation  to quantum processes, since it is  not only a fundamental equation in the theory of stochastic processes but also a necessary condition for a stochastic process to be specified as Markovian~\cite{Gardiner2009,Kampen2007}.
	\par  Motivated by this fundamental issue, we first unify the time evolution of the one-point probability vector of stochastic processes and the quantum processes for $N$-level systems by constructing a quantum operational representation of stochastic matrices. The construction is unique with a minimal number of $N$  Kraus operators. This unification allows us to consider the time evolution of stochastic and quantum processes on the same theoretical ground within finite-dimensional spaces. Accordingly, quantum states and quantum operations can be considered, respectively, an extension of the one-point probability vectors and stochastic matrices. Secondly, based on the unification, we prove that CP-divisibility for quantum operations is the natural and consistent extension of the Chapman-Kolmogorov equation. Furthermore, CP-divisibility can hence be demonstrated as a necessary but insufficient condition for a quantum process to be specified as Markovian. 
	\par The paper is organized as follows. We present some fundamental concepts of stochastic and quantum processes in Sections \ref{subsection: introduction stochastic process} and \ref{subsection: introduction quantum operations}, respectively. After establishing the quantum operational representation of stochastic matrices in Section \ref{sec:quantum operational representation}, we prove in Section \ref{sec:extension of P-divibility} that CP-divisibility is the extension of the Chapman-Kolmogorov equation.  We illustrate the theoretical results in Section \ref{sec: application} by application to a dichotomic Markov process, and finally discuss the results in Section \ref{sec:conclusion}. 
	\subsection{Classical P-divisible processes}\label{subsection: introduction stochastic process}
	A stochastic process $X_{t}$ is called a Markov process if the corresponding conditional probabilities satisfy
	\begin{equation}\label{Markov condition}
		\begin{split}
			p_{1\mid k}(x_{k+1},t_{k+1}\mid x_{k},t_{k};\ldots;x_{2},t_{2};x_{1},t_{1})
			= p_{1\mid 1}(x_{k+1},t_{k+1}\mid x_{k},t_{k})    
		\end{split}
	\end{equation}
	for all hierarchies of any order $k$ and the ordered time instants $t_{1}\leq t_{2}\leq \ldots \leq t_{k}\leq t_{k+1}$. We use the notation $p(x_{k+1},t_{k+1}\mid x_{k},t_{k}):=p_{1\mid 1}(x_{k+1},t_{k+1}\mid x_{k},t_{k})$ and the \textit{one-point probability} $p(x_{k},t_{k}):=p_{1}(x_{k},t_{k})$  from here on for simplicity. 
	
	The time evolution of stochastic processes has been developed based on the time evolution of the \textit{one-point probability distribution} \cite{Feller1970V2, Gardiner2009, Kampen2007} which, for the processes having finite sample space, is expressed as
	\begin{equation}\label{First Markov evolution equation}
		p(x,t)=\sum_{x^{\prime}}T(x,t\mid x^{\prime},t^{\prime})p(x^{\prime},t^{\prime})
	\end{equation}
	where $T(x,t\mid x^{\prime},t^{\prime}):=p(x,t\mid x^{\prime},t^{\prime})$. Equation (\ref{First Markov evolution equation}) connects the probability distribution $p(x,t)$ at time $t$ to the probability distribution $p(x^{\prime},t^{\prime})$ at an earlier time $t^{\prime}$ by means of the transition (or conditional) probabilities $T(x,t\mid x^{\prime},t^{\prime})$. Considering the finite sample space $\mathbb{E}=\{x_{1},x_{2},\ldots,x_{N}\}$, and the convex cone of the column probability vectors $\mathbb{P}^{N}=\{(p_{1},p_{2},\ldots,p_{N})\in \mathbb{R}^{N}\mid \forall p_{i}\geq 0, \sum_{i=1}^{N}p_{i}=1\}$, equation (\ref{First Markov evolution equation}) can be expressed in a compact form as
	\begin{equation}\label{Second Markov evolution equation}
		\mathbf{p}(t)=T(t, t^{\prime})\mathbf{p}(t^{\prime})
	\end{equation}
	such that $\mathbf{p}(s)=(p(x_{1},s),p(x_{2},s),\ldots,p(x_{N},t))\in\mathbb{P}^{N}$ for $s\in\{t,t^{\prime}\}$ and  $T(t, t^{\prime})$ is the transition matrix with $T_{jk}(t,t^{\prime})=T(x_{j},t\mid x_{k},t^{\prime})$.  A general transition matrix $T(t,t')$ is characterized  by the following properties: 
	\begin{enumerate}
		\item [1.] Its elements are non-negative, $T_{jk}(t,t')\geq 0$,
		\item [2.] $\sum_{j=1}^{N}  T_{jk}(t,t')=1$ for all $k$,
		\item [3.] $T_{jk}(t',t')=\delta_{j,k}$,
		\item [4.] $p_2(x_{j},t;x_{k},t')=T_{jk}(t \mid t')p(x_{k},t')$ for all $t \geq t'$.
	\end{enumerate}  
	 The transition matrices of Markov processes satisfy the celebrated Chapman-Kolmogorov equation,
	\begin{equation}\label{Chapman-Kolmogorov for transition matrix}
		T(t,t^{\prime})=T(t,s)T(s,t^{\prime}), \quad t\geq s\geq t^{\prime}.
	\end{equation}
	If the fourth property is dropped,  the notion of a transition matrix, $T(t,t_{1})$, generalizes to that of a stochastic matrix $\Lambda(t,t_{1})$, which is no longer constrained by property 4.  Consequently, equation (\ref{Second Markov evolution equation}) for any classical stochastic process can be recast in the following form:
	\begin{equation}\label{One point time evolution}
		\mathbf{p}(t)=\Lambda(t,t_{1})\mathbf{p}(t_{1})
	\end{equation}
	where $t_{1}$ is the initial time.   We note that $\lim_{t\rightarrow t_{1}}\Lambda(t,t_{1})=I_{N}$ must hold to maintain consistency. Assuming that $\Lambda_{jk}(t,t_{1})=\lambda_{jk}$, equation (\ref{One point time evolution}) takes the following form
	\begin{equation}
		p_{j}(t)=\sum_{k=1}^{N}\lambda_{jk}p_{k}(t_{1})    
	\end{equation}
	with $p_{l}(s)=p(x_{l},s)$.  A stochastic matrix $\Lambda(t_{1})$ is \textit{divisible} if, for any $t\geq s\geq t_{1}$, one can write
	\begin{equation}
		\Lambda (t,t_{1})=\Lambda(t,s)\Lambda(s,t_{1}).
	\end{equation}
\begin{sloppypar}
	If $\Lambda(s,t_{1})$ is invertible,  $\Lambda(t,s)$ is uniquely determined as $\Lambda(t,s)=\Lambda(t,t_{1})\Lambda^{-1}(s,t_{1})$. It should be noted that $\Lambda(t,s)$ does not have to be a stochastic matrix since the elements of $\Lambda(t,s)$ might be negative.  P-divisibility for stochastic processes having finite sample space characterized by $\Lambda (t,t_{1})$ is defined as follows\cite{Rivas2014a, CanturkBreuer2024}:
\end{sloppypar}	
	\begin{definition}\label{first_definition}
		A stochastic process $X_{t}$ is called 
		\textit{positively divisible (classical P-divisible)} if $\Lambda(t,t_{1})=\Lambda(t,s)\Lambda(s,t_{1})$ such that 
		$\Lambda(t,s)$ is also a stochastic matrix for all $t\geq s\geq t_{1}$.
	\end{definition}
	We emphasize that classical P-divisibility is the extension of the Chapman-Kolmogorov equation when the notion of the transition matrix is generalized to that of the stochastic matrix. Conversely, classical P-divisibility is uniquely reduced to the Chapman-Kolmogorov equation whenever the stochastic matrix of any stochastic process is equivalent to the transition matrix \cite{CanturkBreuer2024, Vacchini-Smirne-Breuer2011}. Markov processes are classical P-divisible. However, there are some classical P-divisible non-Markovian processes (see  \ref{example for non-Markovian process}). Therefore, we state the following remark for our purpose:  
	\begin{remark}
		In general, classical P-divisibility is a necessary but insufficient condition for a stochastic process to be classified as Markovian.  
	\end{remark}
	\subsection{Quantum operations }\label{subsection: introduction quantum operations}
	Let us assume an open quantum system within finite $N$-dimensional Hilbert space $\mathcal{H}^{N}$ and with the generic quantum state $\rho\in\mathcal{B}(\mathcal{H}^{N})\subset \mathcal{M}_{N}(\mathbb{C})$.  Quantum dynamics of the  system is represented by a completely positive and trace-preserving ($CPTP$) linear map,  $\Phi(t,t_{1}):\mathcal{B}(\mathcal{H}^{N})\rightarrow \mathcal{B}(\mathcal{H}^{N})$ such that the quantum state $\rho(t)$ of the system evolving from the initial state $\rho(t_{1})$ is given by
	\begin{equation}
		\rho(t)=\Phi(t,t_{1})\rho(t_{1}).
	\end{equation}
	The $CPTP$ map $\Phi(t,t_{1})$ is called a \textit{quantum operation}~\cite{Nielsen-Chuang2010} and admits a Kraus representation 
	\begin{equation}
		\Phi(t,t_{1})\rho(t_{1})=\sum_{k=1}^{M}A_{k}(t,t_{1})\rho(t_{1})A^{\dagger}_{k}(t,t_{1}),
	\end{equation}
	with $A_{k}$ being Kraus operators and $M\leq N^{2}$ \cite{Watrous2018}. Moreover, $\lim_{t\rightarrow t_{1}}\Phi(t,t_{1})=I_{N}$. A quantum operation $\Phi(t,t_{1})$ is called \textit{divisible} if 
	\begin{equation}\label{quatum operation divisible}
		\Phi(t,t_{1})=\Phi(t,s)\circ\Phi(s,t_{1}) 
	\end{equation}
	for all $t\geq s\geq t_{1}$ such that $\Phi(t,s)$ does not need to be a quantum operation. In addition, $\Phi(t,t_{1})$ is called \textit{positively divisible} (P-divisible) if $\Phi(t,s)$ in equation (\ref{quatum operation divisible}) is a positive trace-preserving map, and \textit{completely positive divisible} (CP-divisible) if $\Phi(t,s)$ in equation (\ref{quatum operation divisible}) is also a quantum operation, that is, a $CPTP$ map which admits a Kraus representation. 
	\par The \textit{matrix form} of a quantum operation $\Phi$, which is also known as the \textit{natural representation} \cite{Watrous2018}, is defined in terms of the corresponding Kraus operators $\{A_{k},k=1,2,\ldots,M\}$ as follows: 
	\begin{equation*}
	\ensuremath{	M[\Phi]:=\sum_{k=1}^{M}A_{k}\otimes \overline{A}_{k}},
	\end{equation*} 
	where $\overline{A}_{k}$ is the complex conjugate of $A_{k}$. We note that $M[\Phi]$ is a matrix acting on the Hilbert space $\mathcal{H}^{N}\otimes\mathcal{H}^{N}$. The action of a quantum operation $\Phi$ on a quantum state $\rho$ is equivalent to the matrix product of its matrix form with the vector $\mathrm{vec}(\rho)$: $\Phi (\rho)\equiv M[\Phi]\mathrm{vec}(\rho)$. The map $\mathrm{vec}$ is defined as follows. Let $A=(a_{jk})\in \mathcal{M}_{K\times\letsymb{N}}(\mathbb{C})$ be a general $K\times\letsymb{N}$ matrix, and $\{\ket{f_{j}},\; j=1,2,\ldots, K\}$ and $\{\ket{e_{k}},\; j=1,2,\ldots, N\}$ be the standard bases of the Hilbert spaces $\mathcal{H}^{K}$ and $\mathcal{H}^{N}$ respectively. Then $A=\sum_{j,k}a_{jk}\ket{f_{j}}\bra{e_{k}}$. The map $\mathrm{vec}:\mathcal{M}_{K\times\letsymb{N}}(\mathbb{C})\rightarrow \mathcal{H}^{K}\otimes \mathcal{H}^{N}$ is defined as \[\mathrm{vec}(A):=\sum_{j,k}a_{jk}\ket{e_{j}}\otimes\ket{f_{k}}.\] 
	\begin{definition}
		Let $\Phi_{1}$ and $\Phi_{2}$ be two quantum operations acting on $\mathcal{B}(\mathcal{H}^{N})$ with the respective sets of the Kraus operators, $\{A_{k}, \; k=1,2,\ldots, M\leq N^{2}\}$ and $\{B_{j},\; j=1,2,\ldots, K\leq N^{2}\}$. Then, $\Phi_{1}$ and $\Phi_{2}$ are said to be essentially the same if their matrix forms are the same. 
	\end{definition}
	We note that two quantum operations are essentially the same iff there exists an isometry $T=(t_{jk})_{K\times\letsymb{M} }$ such that $B_{j}=\sum_{k}t_{jk}A_{k}$. In that case, $T^{\dagger}T=I_{M}$ (see in particular the second chapter of Ref. \cite{Watrous2018}).
	\par Finally, we wish to introduce some concepts that will be used. The \textit{Hadamard product} of two matrices $A=(a_{jk}),\, B=(b_{jk})\in\mathcal{M}_{K\times\letsymb{N}}(\mathbb{C})$ is defined as $A\odot\letsymb{B}:=(a_{jk}b_{jk})\in\mathcal{M}_{K\times\letsymb{N}}(\mathbb{C})$, and the \textit{Hadamard power} of a matrix as $A^{\odot\letsymb{r}}=(a_{jk}^{r})\in\mathcal{M}_{K\times\letsymb{N}}(\mathbb{C})$ for  $r\in\mathbb{R}$ \cite{GarciaHorn2017, HornJohnson2013}. 
	\section{Quantum operational representation of stochastic matrices}\label{sec:quantum operational representation}
	Let us consider  a fixed orthonormal basis $B_{1}=\{\ket{e_{k}},k=0,\ldots, N-1\}$ for $\mathbb{R}^{N}$. Then, a probability vector $\mathbf{p}(t)\in\mathbb{P}^{N}$ can be written as
	\begin{equation}
		\mathbf{p}(t)=\sum_{k=0}^{N-1}p(x_{k},t)\ket{e_{k}}
	\end{equation}
	for all $t\geq t_{1}$. Also, consider the convex cone of the diagonal quantum states, $\mathcal{D}(\mathcal{H}^{N})=\{\rho\in\mathcal{B}(\mathcal{H}^{N})\mid \rho_{jk}=\rho_{kk}\delta_{jk}\}$ and the corresponding fixed 
	orthonormal basis $\Tilde{B}_{1}=\{\ket{f_{k}}\bra{f_{k}}, k=0,\ldots, N-1\}$, which might be the same as $B_{1}$. $\Tilde{B}_{2}=\{\ket{f_{j}}\bra{f_{k}},\; j,k=0,\ldots, N-1\}$ denotes the orthonormal basis for the set of the quantum states $\mathcal{B}(\mathcal{H}^{N})$ so that $\Tilde{B}_{1}\subset\Tilde{B}_{2}$ . The linear bijective map $\mathit{F}$ is defined as $\mathit{F}:\mathbb{P}^{N}\longrightarrow \mathcal{D}(\mathcal{H}^{N})$ with $\mathit{F}\mathbf{p}(t)=\rho(t)\in \mathcal{D}(\mathcal{H}^{N})$ such that 
	\begin{equation*}
	\ensuremath{	\thinmuskip=0mu \mathit{F}\mathbf{p}(t):=\sum_{k=0}^{N-1}p(x_{k},t) \mathit{F}(\ket{e_{k}})
		= \sum_{k=0}^{N-1}p(x_{k},t) \ket{f_{k}}\bra{f_{k}}  }
	\end{equation*}
	for all $t\geq t_{1}$. If a probability vector $\mathbf{p}(t_{1})$ evolves under the stochastic matrix $\Lambda(t,t_{1})=(\lambda_{jk})_{N\times\letsymb{N}}$ to   $\mathbf{p}(t)$, then $\mathbf{p}(t)=\Lambda(t,t_{1})\mathbf{p}(t_{1})$ and accordingly
	\begin{equation*}
		\ensuremath{\begin{aligned}
		\rho(t)&=\mathit{F}\mathbf{p}(t)=\mathit{F}(\Lambda(t,t_{1})\mathbf{p}(t_{t})) \\ 
			&=\sum_{j,k=0}^{N-1}\lambda_{jk}p(x_{k},t_{1})\ket{f_{j}}\bra{f_{j}}
			=\sum_{j=0}^{N-1}p(x_{j},t)\ket{f_{j}}\bra{f_{j}}.
		\end{aligned}}
	\end{equation*}
	Also, the diagonalization operation $ \Pi(\rho(t))$ on $\mathcal{B}(\mathcal{H}^{N})$ is defined  as
	\begin{equation}\label{diagonalization process}
		\Pi(\rho):=\sum_{k=0}^{N-1}\ket{f_{k}}\bra{f_{k}}\rho\ket{f_{k}}\bra{f_{k}}, \quad \rho \in \mathcal{B}(\mathcal{H}^{N}).
	\end{equation}
	\par
\begin{figure}
		\centering
		\includegraphics{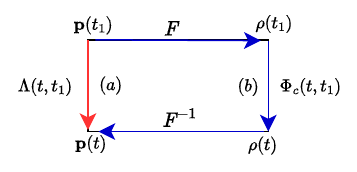}
		\caption{$(a)$ (red): The time evolution of the initial probability vector $\mathbf{p}(t_{1})$ under the stochastic matrix $\Lambda(t,t_{1})$ to the final probability vector $\mathbf{p}(t)$ is equivalent to the operation in $(b)$ (blue), which states that $\mathbf{p}(t_{1})$ is first mapped by $\mathit{F}$ to the quantum state $\rho(t_{1})$ and then, $\rho(t_{1})$ evolves  under the quantum operation $\Phi_{c}(t,t_{1})$ to the diagonal quantum state $\rho(t)$, and finally, the probability vector corresponding to $\rho(t)$ under the action of $\mathit{F}^{-1}$ is the final probability vector $\mathbf{p}(t)$.}
		\label{figure 2}
	\end{figure}

 We now construct a quantum operation $\Phi_{c}(t,t_{1})$  that represents a stochastic matrix $\Lambda(t,t_{1})$. The aforementioned quantum operation  $\Phi_{c}(t,t_{1})$  (see Fig. \ref{figure 2}) satisfies the following relation 
	\begin{equation}\label{stochastic to quantum representation}
		\Lambda(t,t_{1})\mathbf{p}(t_{1})=\mathit{F}^{-1}\circ \Phi_{c}(t,t_{1})\circ \mathit{F}\mathbf{p}(t_{1})    
	\end{equation}  
	for all $\mathbf{p}(t_{1})\in\mathbb{P}^{N}$ and $t\geq s \geq t_{1}$ as well as the following characteristic properties:
	\begin{enumerate}
		\item[(i)] $\Phi_{c}(t,t_{1})\rho(t_{1})\in\mathcal{D}(\mathcal{H}^{N})$ for all $\rho(t_{1})\in\mathcal{D}(\mathcal{H}^{N})$.
		\item[(ii)] \begin{sloppypar}
			$\Pi(\Phi_{c}(t,t_1)\rho(t_{1}))=\Phi_{c}(t,t_{1})\Pi(\rho(t_{1}))$ for all $\rho(t_{1})\in\mathcal{B}(\mathcal{H}^{N})$.
		\end{sloppypar} 
		\item[(iii)] (\textit{Stability Condition}) If the stochastic matrix $\Lambda (t,t_{1})$ for $0\leq t<\infty$ is invertible, $\Phi_{c}(t,t_1)$ is also invertible. 
	\end{enumerate}

	$\Phi_{c}(t,t_{1})$ is called a \textit{quantum operational representation} which is \textit{formally valid} in every finite dimension.  Here, the quantum states $\rho(t)\in\mathcal{B}(\mathcal{H}^{N})$ are given in the orthonormal basis $\Tilde{B}_{2}$.   Note that the injectivity of the map $\Lambda(t,t_{1})\rightarrow \Phi_{c}(t,t_{1})$ is ensured through equation (\ref{stochastic to quantum representation}). The first characteristic property is necessary for equation (\ref{stochastic to quantum representation}) to be satisfied. The second characteristic property states that the quantum operational representation $\Phi_{c}(t,t_{1})$ does not have any effect on the diagonal elements of the resultant quantum state which is not contained in $\Lambda(t,t_{1})$.  In other words, the representation per se cannot induce any quantum effect on the diagonal elements of the resultant quantum state.  The third property ensures  the existence of the reverse of the blue arrows (labeled by \textit{(b)}) when the red arrow (labeled by \textit{(a)}) is reversed in Fig.\ref{figure 2}. This property is required for the stability of the representation, that is, a small reversible change in the stochastic matrix should not result in an irreversible change in the quantum operational representation. For instance, if, after a small change, the stochastic matrix is still invertible, the quantum operational representation should also remain invertible. We point out that equation (\ref{stochastic to quantum representation}), with its characteristic properties $\mbox{(i)-(iii)}$, is the \textit{embedding} of stochastic matrices into quantum operations. 
	\par We note that the algebraic approach to stochastic processes developed by Accardi et. al.~\cite{Accardi1982} should not be confused with the quantum operational representation presented here. The authors in their approach  generalized the \textit{"notions of 'random variable' and 'stochastic process' by stating them in a purely algebraic way"}~\cite{Fagnola1999} so that they can be applied successfully to any measurable space. Thereby, one can obtain a stochastic process.  However, the quantum operational representation presented above is an embedding that is contingent upon the physically contentful characteristic properties.    
	
	\begin{theorem}{(Existence Theorem)}\label{theorem1}
		There exists a quantum operational representation $\Phi_{c}(t,t_{1})$ of a general stochastic matrix $\Lambda(t,t_{1})$ acting on $\mathbb{R}^{N}$ whose Kraus representation is given by the Kraus operators 
		\begin{equation}\label{Kraus operators}
			A_{s}(t,t_{1})=(a_{jk}^{(s)})_{N\times\letsymb{N}}, \quad a_{jk}^{(s)}:=\frac{\sqrt{\lambda_{jk}}}{\sqrt{N}}\e^{\frac{2\pi \mathrm{i}s(j-k)}{N}}
		\end{equation}
		with $s,j,k\in\{0,1,\ldots,N-1\}$ and $\lambda_{jk}=\Lambda_{jk}(t,t_{1})$.
	\end{theorem}
	\begin{proof}
		The time dependence of the Kraus operators is implicitly given in the elements  $\lambda_{jk}$ of the stochastic matrix $\Lambda(t,t_{1})$. Let us first note that for any integer number $\alpha$ we have the following well-known identity   
		\begin{equation}\label{number theory identity}
			\sum_{s=0}^{N-1}\frac{\e^{\frac{2\pi\mathrm{i}\alpha s}{N}}}{N}\begin{cases}
				=0& \mbox{if}\quad 0<\alpha\mod(N)\leq N-1,\\
				=1              & \mbox{if}\quad  \alpha\mod(N)=0.
			\end{cases}
		\end{equation}
		Then, one can directly check that the Kraus operators in equation (\ref{Kraus operators}) satisfy the identity condition, $\sum_{s=0}^{N-1}A_{s}^{\dagger}A_{s}=\mathit{I}_{N}$.  Now, for a general initial probability vector $\mathbf{p}(t_{1})=\sum_{m=0}^{N-1}p(x_{m},t_{1})\ket{e_{m}}$ and the quantum operation $\Phi_{c}(t,t_{1})$ with the Kraus operators given by equation (\ref{Kraus operators}), equation (\ref{stochastic to quantum representation}) yields
		\begin{subequations}
			\begin{equation}\label{Equation for second characteristic condition}
				\begin{split}
					&(\Phi(t,t_{1})\circ \mathit{F}\mathbf{p}(t_{1}))_{jk}=(\Phi(t,t_{1})\rho(t_{1}))_{jk} =\sum_{s,m,l=0}^{N-1}(A_{s})_{jm}\rho(t_{1})_{ml}(\overline{A}_{s})_{kl}\\
					& \quad =\sum_{s,m,l=0}^{N-1}p(x_{m},t_{1})\sqrt{\lambda_{jm}\lambda_{kl}}\frac{\delta_{ml}}{N}\e^{\frac{2\pi\mathrm{i}s(j+l-m-k)}{N}},
				\end{split}       
			\end{equation}
			\text{so that using equation (\ref{number theory identity}), we obtain}
			\begin{equation}
				\begin{split}
					&(\Phi(t,t_{1})\circ \mathit{F}\mathbf{p}(t_{1}))_{jk}=\sum_{m=0}^{N-1}p(x_{m},t_{1})\sqrt{\lambda_{jm}\lambda_{km}}\left(\sum_{s=0}^{N-1}\frac{1}{N}\e^{\frac{2\pi\mathrm{i}s(j-k)}{N}}\right)\\
					&\quad =\sum_{m=0}^{N-1}p(x_{m},t_{1})\sqrt{\lambda_{jm}\lambda_{km}}\delta_{jk} =\sum_{m=0}^{N-1}\lambda_{jm}p(x_{m},t_{1})\delta_{jk},
				\end{split}
			\end{equation}   
		\end{subequations}
		which is a function of $t$ and $\delta_{jk}$ is the Kronecker delta. Defining 
		\begin{equation}
			\rho(t)_{jk}:=\sum_{m=0}^{N-1}\lambda_{jm}p(x_{m},t_{1})\delta_{jk}    
		\end{equation} 
		we see that $\rho(t)\in\mathcal{D}(\mathcal{H}^{N})$. Hence, the characteristic property \mbox{(i)} is satisfied. Furthermore, 
		\begin{equation}
			\begin{split}
				\mathit{F}^{-1}(\rho(t))&=\sum_{m=0}^{N-1}\lambda_{jm}p(x_{m},t_{1})\mathit{F}^{-1}(\ket{f_{j}}\bra{f_{j}})\\
				&=\sum_{m=0}^{N-1}\lambda_{jm}p(x_{m},t_{1})\ket{e_{j}}=\Lambda(t,t_{1})\mathbf{p}(t_{1}),  
			\end{split}
		\end{equation}
		so that equation (\ref{stochastic to quantum representation}) is satisfied. Based on equation (\ref{Equation for second characteristic condition}), one can directly check that the characteristic property \mbox{(ii)} is satisfied.   The characteristic property \mbox{(iii)} is also satisfied, as we will prove in  Lemma \ref{lemma quantum operation inverse} below.
	\end{proof}
	\par In order to present Lemma \ref{lemma quantum operation inverse}, we firstly demonstrate the matrix form of the quantum operational representation $\Phi_{c}(t,t_{1})$ of Theorem \ref{theorem1}. To this end, it is evident that the Kraus operators $\{A_{s}\}_{s=0}^{N-1}$ are the Hadamard product of two matrices: $A_{s}=\Lambda(t,t_{1})^{\odot 1/2 }\odot U_{s}$, where the entries of the matrix $U_{s}$  are $u^{(s)}_{jk}=\frac{1}{\sqrt{N}}exp(\frac{2\pi\mathrm{i}s(j-k)}{N})$. Then, the matrix form $M_{c}(t,t_{1})$ of the quantum operation $\Phi_{c}(t,t_{1})$ is equal to
	\begin{equation}\label{matrix form general}
		\begin{split}
			M_{c}(t,t_{1})&=\displaystyle\sum_{s=0}^{N-1}A_{s}\otimes\overline{A}_{s}\\
			&=\displaystyle\sum_{s=0}^{N-1}(\Lambda^{\odot 1/2}(t,t_{1})\odot U_{s})\otimes(\Lambda^{\odot 1/2}(t,t_{1})\odot \overline{U}_{s})\\
			&=\displaystyle\sum_{s=0}^{N-1}(\Lambda^{\odot 1/2}(t,t_{1})\otimes\Lambda^{\odot 1/2}(t,t_{1}))\odot(U_{s}\otimes\overline{U}_{s})\\
			&=(\Lambda^{\odot 1/2}(t,t_{1})\otimes\Lambda^{\odot 1/2}(t,t_{1}))\odot\displaystyle\sum_{s=0}^{N-1}(U_{s}\otimes\overline{U}_{s}),
		\end{split} 
	\end{equation}
	where we have used the equality $(A\odot C)\otimes (B\odot D)=(A\otimes B)\odot(C\otimes D)$ in the second line.  Using equation (\ref{number theory identity}), it is straightforward to show that, for $j,k,l,m=0,1,\ldots,N-1,$ $({G}_{N})_{(N.j+k)(N.l+m)}:=$$\sum_{s=0}^{N-1}(U_{s}\otimes\overline{U}_{s})_{(N.j+k)(N.l+m)}$$=\sum_{s=0}^{N-1}(U_{s})_{jl}(\overline{U}_{s})_{km}=\delta_{0,(j-k+m-l)\mbox{mod}(N)}$ which yields the block matrix form of  $G_{N}$ as
	\begin{equation}
		G_{N}=\begin{pmatrix}
			\mathit{I}_{N} & C_{N} & C_{N}^{2}&\cdots&C_{N}^{N-1}\\
			C_{N}^{N-1} &\mathit{I}_{N}& C_{N}&\cdots &C_{N}^{N-2}\\
			C_{N}^{N-2}&C_{N}^{N-1}&\mathit{I}_{N}&\cdots &C_{N}^{N-3}\\
			\vdots&\vdots &\vdots & &\vdots \\
			C_{N} & C_{N}^{2}&C_{N}^{3}&\cdots & \mathit{I}_{N}
		\end{pmatrix}.
	\end{equation}
	Here, $C_{N} =(c_{jk})_{N\times\letsymb{N}}:=(\delta_{(j+1)\mbox{mod}(N),k})_{N\times\letsymb{N}}$ is the $N\times\letsymb{N}$ \textit{basic circulant permutation} matrix having the property $C_{N}^{N}=I_{N}$ \cite{HornJohnson2013}. We note that $G_{N}$ is the block circulant matrix of the elements $\{\mathit{I}_{N},C_{N},C_{N}^{2},\ldots,C_{N}^{N-1}\}$.  Moreover, noting that $G_{N}^{\dagger}=G_{N}$ and $G^{2}_{N}=NG_{N}$, $G_{N}$ is a \textit{positive semidefinite} matrix satisfying $G_{N}= B^{\dagger}B$ where $B= (1/\sqrt{N})G_{N}$.  $M_{c}(t,t_{1})$ can be written as
	\begin{equation}\label{matrix representation of quantum operation for stochastic matrices}
		M_{c}(t,t_{1})=(\Lambda^{\odot 1/2}\otimes\Lambda^{\odot 1/2})\odot G_{N},
	\end{equation}
	where the time dependence of $\Lambda(t,t_{1})$ has been omitted for simplicity. The structure of $G_{N}$ allows us to partition  $M_{c}(t,t_{1})$ into $N$ principal submatrices, each of which is determined by $N$ certain rows (or columns) whose possible nonzero entries are at the same positions. Interestingly, $M_{c}(t,t_{1})$ then becomes the direct sum of these principal submatrices.  For instance, for $N=2$ and the stochastic matrix $\Lambda(t,t_{1})=(\lambda_{jk})_{2\times\letsymb{2}}$, we have 
	
	\begin{equation}\label{two dimensional representation matrix}
		M_{c}(t,t_{1})=\begin{pmatrix}
			\lambda_{00} & 0 & 0 & \lambda_{01}\\
			0 & \sqrt{\lambda_{00}\lambda_{11}} & \sqrt{\lambda_{01}\lambda_{10}} & 0\\
			0 & \sqrt{\lambda_{10}\lambda_{01}} &\sqrt{\lambda_{11}\lambda_{00}} & 0\\
			\lambda_{10} & 0 & 0 & \lambda_{11}
		\end{pmatrix}
	\end{equation}
	such that its rows can be partitioned into two sets with the first and fourth rows being in one set and the second and third in the other.   One can recognize that the partition is systematically obtained  based on the following rules: 
	\begin{enumerate}
		\item [1.] Group the index set $\{0,1,\ldots,N^{2}-1\}$ into the disjoint subsets $n_{j}=\{jN+0,jN+1,\ldots,jN+N-1\}$ for $j=0,1,\ldots,N-1$.
		\item[2.] Define the disjoint subsets $\alpha_{j}$  as follows:  $\alpha_{j}=\{n_{0}(j\mbox{mod}(N)),n_{1}((j+1)\mbox{mod}(N)),\ldots,n_{N-1}((j+N-1)\mbox{mod}(N))\}$ for $j=0,1,\ldots, N-1$ with $n_{k}((j+k)\mbox{mod}(N))$ being the $((j+k)\mbox{mod}(N))^{th}$ element of $n_{k}$.
		\item[3.] Then, 
		$M_{c}(t,t_{1})=M_{c}(t,t_{1})[\alpha_{0}]\oplus\letsymb{M}_{c}(t,t_{1})[\alpha_{1}]\oplus\cdots\oplus\letsymb{M}_{c}(t,t_{1})[\alpha_{(N-1)}]$, where $ M_{c}(t,t_{1})[\alpha_{j}]$ is the principal submatrix of entries that lie in the rows and columns of $M_{c}(t,t_{1})$ indexed by the set $\alpha_{j}$. In particular, $M_{c}(t,t_{1})[\alpha_{0}]=\Lambda(t,t_{1})$. 
	\end{enumerate}
	Applying these rules, for example, to the matrix form in equation (\ref{two dimensional representation matrix}), we first obtain the following index subsets: $\alpha_{0}=\{0,3\}$, $\alpha_{1}=\{1,2\}$. Accordingly, $M_{c}(t,t_{1})$ can be written as $M_{c}(t,t_{1})=V_{0}\oplus\letsymb{V}_{1}$, where $V_{j}=M_{c}(t,t_{1})[\alpha_{j}]$ for $j=0,1$, and in particular, $V_{0}=\Lambda(t,t_{1})$.
	\par One can further see that, taking $\Lambda(t,t_{1})$ as $\Lambda$ for simplicity, the principal submatrices $\{M_{c}(t,t_{1})[\alpha_{j}],\,j=0,1,\ldots, N-1\}$ have the compact form $M_{c}(t,t_{1})[\alpha_{j}]=\Lambda^{\odot\letsymb{1/2}}\odot(\letsymb{(C_{N}^{j})^{T}}\Lambda^{\odot\letsymb{1/2}}\letsymb{C_{N}^{j}})$ with $(C_{N}^{j})^{T}$ being the transpose of $C_{N}^{j}$. For example, considering the matrix form $M_{c}(t,t_{1})$ in equation (\ref{two dimensional representation matrix}) again, $M_{c}(t,t_{1})[\alpha_{0}]=\Lambda=\Lambda^{\odot\letsymb{1/2}}\odot\Lambda^{\odot\letsymb{1/2}}$, and $M_{c}(t,t_{1})[\alpha_{1}]=\Lambda^{\odot\letsymb{1/2}}\odot(\letsymb{C_{N}^{T}}\Lambda^{\odot\letsymb{1/2}}\letsymb{C_{N}})$. We sum up all of these observations in the following remark:
	\begin{remark}\label{formula of vj}
		The matrix form $M_{c}(t,t_{1})$ of the quantum operational representation $\Phi_{c}(t,t_{1})$ can be expressed in the following compact form: 
		\begin{equation}\label{principal submatrix decomposition of M}
			M_{c}(t,t_{1})=V_{0}\oplus\letsymb{V}_{1}\oplus\cdots\oplus\letsymb{V}_{N-1},
		\end{equation}
		such that $V_{j}=\Lambda^{\odot\letsymb{1/2}}\odot((C_{N}^{j})^{T}\Lambda^{\odot\letsymb{1/2}}C_{N}^{j})$ for $j=0,1,\ldots,N-1$. In particular, $V_{0}=\Lambda=\Lambda(t,t_{1})$. 
	\end{remark}
	It is noteworthy that if the stochastic matrix $\Lambda$ is a circulant matrix, then $V_{0}=V_{1}=\cdots=V_{N-1}=\Lambda$. This is so, since for any circulant matrix $A\in\mathcal{M}_{N}(\mathbb{C})$, the equality $C_{N}^{T}AC_{N}=A$ holds. We note that if the stochastic matrix $\Lambda$ is circulant, then $\Lambda^{\odot\letsymb{1/2}}$ is also circulant.   
	\begin{lemma}\label{lemma quantum operation inverse}
		If the stochastic matrix $\Lambda(t,t_{1})$ is invertible for $0\leq t<\infty$, its quantum operational representation $\Phi_{c}(t,t_{1})$ is also invertible.
	\end{lemma}
	\begin{proof}
		Let us take $t_{1}=0$ for simplicity and consider $\Lambda(t,0)$ be invertible for $0\leq t<\infty$.  One must now show that each principal submatrix in equation (\ref{principal submatrix decomposition of M}) is invertible. For infinitesimal time, $t=\varepsilon<<1$, $\Lambda(\varepsilon,0)= I+\varepsilon\gamma\letsymb{W}+O(\varepsilon^{2})$, where $\gamma\letsymb{W}=\frac{d\letsymb{\Lambda}(t,0)}{d\letsymb{t}}(t=0)$ and the positive constant $\gamma$ is the relaxation rate of the underlying stochastic process. Accordingly, $V_{j}(\varepsilon,0)=\letsymb{I}+\varepsilon\gamma\letsymb{\tilde{W}}+O(\varepsilon^{2})$ such that $\tilde{W}_{kk}=\frac{1}{2}(W_{kk}+((C_{N}^{j})^{T}WC_{N}^{j})_{kk})$ and $\tilde{W}_{kl}=(W^{\odot\letsymb{1/2}}\odot((C_{N}^{j})^{T}W^{\odot\letsymb{1/2}}C_{N}^{j}))_{kl}$ for $k\neq\letsymb{l}$. Then, the determinant of $\Lambda(\varepsilon,0)$ is \begin{equation}
			det(\Lambda(\varepsilon,0))= \letsymb{1}+\varepsilon\gamma\tr(W)+O(\varepsilon^{2})\approx\letsymb{1}+\varepsilon\gamma\tr(W)
		\end{equation}
		which is not zero. $\tr(W)$ is the trace of $W$. Similarly, 
		\begin{equation}
			\begin{split}
				det(V_{j}(\varepsilon,0))&=\letsymb{1}+\varepsilon\gamma\tr(\tilde{W})+O(\varepsilon^{2})\approx\letsymb{1}+\varepsilon\gamma\tr(\tilde{W})\\
				&=\letsymb{1}+\varepsilon\gamma\tr(W)\approx\letsymb{det}(\Lambda(\varepsilon,0)).
			\end{split}
		\end{equation} 
		Therefore, the submatrices $V_{j}$ are also invertible. We point out that this result is independent of the relaxation parameter $\gamma$. Hence, the approximation above can also be successfully applied for any time value by choosing a suitable relaxation rate constant $\gamma$ such that $\gamma\letsymb{t}<<1$. 
	\end{proof} 
	\par Quantum operational representation of the stochastic matrices motivates the following definition.  
	\begin{definition}\label{definition3}
		A quantum operation $\Phi(t,t_{1})$ acting on $\mathcal{B}(\mathcal{H}^{N})$ is said to be essentially classic if, for all $t\geq t_{1}$, 
		\begin{enumerate}
			\item [1.] $\Phi(t,t_{1})\rho(t_{1})\in\mathcal{D}(\mathcal{H}^{N})$ for all $\rho(t_{1})\in\mathcal{D}(\mathcal{H}^{N})$, 
			\item[2.]  \begin{sloppypar}
				$\Pi(\Phi(t,t_1)\rho(t_{1}))=\Phi(t,t_{1})\Pi(\rho(t_{1}))$ for all $\rho(t_{1})\in\mathcal{B}(\mathcal{H}^{N})$.
			\end{sloppypar} 
		\end{enumerate}
	\end{definition}
	Here, $\Pi(\cdot)$  is defined as in equation (\ref{diagonalization process}). Note that the properties in Definition \ref{definition3} are exactly the same as the first two characteristic properties of the quantum operational representation $\Phi_{c}(t,t_{1})$.     Interestingly, the structure of the Kraus operators for the essentially classical quantum operations can be uniquely determined in terms of two classes. 
	\begin{theorem}\label{theorem2}
		Essentially classical quantum operations within $N$-dimensional Hilbert space can be uniquely determined in terms of two classes whose Kraus operators have the following forms:
		\begin{description}
			\item[Class 1]\begin{sloppypar}
				It is formed by the sets of quantum operations $C_{1}^{rv}=\{\Psi_{M_{1}}^{(r,v)},\Psi_{M_{2}}^{(r,v)}, \ldots, \Psi_{N^{2}}^{(r,v)};\; 1\leq r,v\leq N+1\}$. Each of the quantum operations $\Psi_{M_{n}}^{(r,v)}$, for fixed values of $r$, $v$ and $M_{n}$, has the set of the Kraus operators $\{R_{s},s=0,1,\ldots,M_{n}-1\}$ such that 
			\end{sloppypar} 
			\begin{equation}\label{true class for quantum operational representation Plus Form}
				R_{s}=\sum_{j,k=0}^{N-1}\frac{\sqrt{\lambda_{jk}}}{\sqrt{M_{n}}}\e^{\frac{2\pi\mathrm{i}s(rj+vk)}{M_{n}}}\ket{f_{j}}\bra{f_{k}},
			\end{equation} 
			where $M_{n}=\max(r,v)(N-1)+n$ with $1\leq n\leq N^{2}-\max(r,v)(N-1)$, and the elements $\lambda_{jk}$ forms an \mbox{$N\times\letsymb{N}$} stochastic matrix $\Lambda$ acting on $\mathbb{P}^{N}$.
			\item[Class 2] \begin{sloppypar}
				It is formed by the sets of quantum operations $C_{2}^{rv}=\{\Phi_{M_{1}}^{(r,v)},\Phi_{M_{2}}^{(r,v)}, \ldots, \Phi_{N^{2}}^{(r,v)};\; 1\leq r,v\leq N+1\}$. Each of the quantum operations $\Phi_{M_{n}}^{(r,v)}$, for fixed values of $r$, $v$ and $M_{n}$, has the set of the Kraus operators $\{R_{s},s=0,1,\ldots,M_{n}-1\}$ such that 
			\end{sloppypar} 
			\begin{equation}\label{true class for quantum operational representation Minus form}
				R_{s}=\sum_{j,k=0}^{N-1}\frac{\sqrt{\lambda_{jk}}}{\sqrt{M_{n}}}\e^{\frac{2\pi\mathrm{i}s(rj-vk)}{M_{n}}}\ket{f_{j}}\bra{f_{k}},
			\end{equation} 
			where $M_{n}=\max(r,v)(N-1)+n$ with $1\leq n\leq N^{2}-\max(r,v)(N-1)$, and the elements $\lambda_{jk}$ forms an \mbox{$N\times\letsymb{N}$} stochastic matrix $\Lambda$ acting on $\mathbb{P}^{N}$.
		\end{description}
	\end{theorem}
	The proof of Theorem \ref{theorem2} has been given in  \ref{Proof_theorem_2}. We point out that the quantum operational representation $\Phi_{c}(t,t_{1})$ is the quantum operation $\Phi_{N}^{(1,1)}$ in the second class of Theorem \ref{theorem2}.
	\par In addition, some quantum operations within a class can be essentially the same.  For example, in $2$-dimensional Hilbert space, $\Phi_{4}^{(3,1)}=\{K_{0}, K_{1},K_{2},K_{3}\}$ and $\Phi_{3}^{(2,1)}=\{R_{0}, R_{1},R_{2}\}$ are essentially the same. Furthermore, we would like to emphasize that certain quantum operations from different classes can also be essentially the same. Partitioning the whole operations in the classes is not addressed in this paper, as it is not relevant to the subject matter.
	\par On the other hand, certain quantum operations in the classes are replicas of one another, suggesting a degree of redundancy. We prove this in the following lemma.
	\begin{lemma}\label{lemma2 redundancy lemma}
		The quantum operations $\{\Psi_{rN}^{(r, r)},\;2\leq\letsymb{r}\leq\letsymb{N}\}$ and $\{\Phi_{rN}^{(r, r)},\;2\leq\letsymb{r}\leq\letsymb{N}\}$ can be obtained by replicating the Kraus operators of the quantum operations $\Psi_{N}^{(1,1)}$ and $\Phi_{N}^{(1,1)}$, respectively. 
	\end{lemma}
	\begin{proof}
		Let the set of the Kraus operators of $\Psi_{rN}^{(r,r)}$ be $\{B_{s}=(b_{jk}^{(s)})_{N\times\letsymb{N}},\; s=0,1,\ldots,rN-1\}$ and that of $\Psi_{N}^{(1,1)}$ be $\{A_{s}=(a_{jk}^{(s)})_{N\times\letsymb{N}},\; s=0,1,\ldots,N-1\}$. Then, in accordance with equations (\ref{true class for quantum operational representation Plus Form}), we have 
		\begin{equation}
			b_{jk}^{(s)}=\frac{\sqrt{\lambda_{jk}}}{\sqrt{r}\sqrt{N}}\e^{\frac{2\pi\mathrm{i}sr(j+k)}{rN}}=\frac{1}{\sqrt{r}}\frac{\sqrt{\lambda_{jk}}}{\sqrt{N}}\e^{\frac{2\pi\mathrm{is(j+k)}}{N}},
		\end{equation}
		which is equal to $(1/\sqrt{r})a_{jk}^{(s)}$ for $s=0,1,\ldots,N-1$. It is then evident that $B_{Nj+s}=(1/\sqrt{r})A_{s}$, which means that $\{B_{s},B_{N+s},\ldots,B_{(r-1)N+s}\}$ are just the replicas of $A_{s}$ multiplied by a number to preserve the identity condition. Following the same reasoning and utilizing equation (\ref{true class for quantum operational representation Minus form}), one can easily conclude the same relation between the couple $\{\{\Phi_{rN}^{(r, r)},\;2\leq\letsymb{r}\leq\letsymb{N}\},\Phi_{N}^{(1,1)}\}$.
	\end{proof}
	\par \begin{sloppypar}
	Lemma \ref{lemma2 redundancy lemma} shows the redundancy of the quantum operations $\{\Psi_{rN}^{(r,r)},\Phi_{rN}^{(r,r)},\;2\leq\letsymb{r}\leq\letsymb{N}\}$. Bearing in mind this fact and utilizing Theorems \ref{theorem1} and \ref{theorem2}, we state the following important result:
	\end{sloppypar}
	\begin{theorem}\label{The first main theorem: uniqueness theorem}{(Uniqueness Theorem)}
		Let $\Lambda(t,t_{1})=(\lambda_{jk})_{N\times\letsymb{N}}$ be a generic stochastic matrix acting on $\mathbb{P}^{N}$, and $2\leq N<\infty$. The quantum operational representation $\Phi_{c}(t,t_{1})$  of the stochastic matrices $\{\Lambda(t,t_{1})\}$ is uniquely determined in the form given by Theorem \ref{theorem1}.
	\end{theorem}
	\begin{proof}
		One should note that the classes of the quantum operations in Theorem \ref{theorem2} consist of all possible quantum operations that satisfy the first two characteristic properties of a quantum operational representation of the generic stochastic matrix $\Lambda(t,t_{1})$. Therefore, it is sufficient to consider only the quantum operations included in the classes to prove the theorem.  The elements $\lambda_{jk}$ of the Kraus operators in the classes are considered to be the entries of $\Lambda(t,t_{1})$.  We show that only the quantum operation $\Phi_{N}^{(1,1)}$, which is equal to the quantum operational representation $\Phi_{c}(t,t_{1})$ of Theorem \ref{theorem1}, satisfies the third characteristic property of quantum operational representation. We outline the proof step by step.
		\par \textit{Step 1.} The form of the quantum operational representation should be valid in every finite dimension, as is explicitly expressed in its definition. Therefore, it is sufficient to show that all quantum operations but $\Phi_{N}^{(1,1)}$ in the classes violate the third characteristic property at least in a particular dimension. We establish this fact in $3$-dimensional Hilbert space. 
		\par \textit{Step 2.} Let us take $N=3$ and consider the  stochastic matrix 
		\begin{equation}\label{dimension three circulant stochastic matrix}
			\Lambda(t,0)=\e^{-\gamma\letsymb{t}}I+2\e^{-\frac{\gamma\letsymb{t}}{2}}A(t,0),
		\end{equation}
		where 
		\begin{equation}
			A(t,0)=\begin{pmatrix}
				a\sinh(\frac{\gamma\letsymb{t}}{2}) &b\sinh(\frac{\gamma\letsymb{t}}{2}) & c\sinh(\frac{\gamma\letsymb{t}}{2})\\
				d\sinh(\frac{\gamma\letsymb{t}}{2}) &  e\sinh(\frac{\gamma\letsymb{t}}{2}) & f\sinh(\frac{\gamma\letsymb{t}}{2})\\
				x\sinh(\frac{\gamma\letsymb{t}}{2}) & y\sinh(\frac{\gamma\letsymb{t}}{2}) & z\sinh(\frac{\gamma\letsymb{t}}{2}) 
			\end{pmatrix}
		\end{equation}
		which can be invertible for $0\!\leq\!\letsymb{t}\!<\!\infty$ for certain values of the nonnegative parameters $\{a,b,c,d,e,f,x,y,z\}$. Note that the parameters satisfy the following equations: (1) $a+d+x=1$, (2) $b+e+y=1$ and (3) $c+f+z=1$.  Below, $\Lambda(t,0)$ is assumed to be the stochastic matrix underlying the quantum operations in the classes. The parameter $\gamma\in\mathbb{R}^{+}$ can be interpreted as the relaxation rate of the relevant process. 
		\par \textit{Step 3.} First of all, one can straightforwardly check through the respective matrix forms that all of the quantum operations $\{\Psi_{M_{n}}^{(r,v)}, \; r,v=1,2,3,4;\; 2\max(r,v)+1\leq\letsymb{M}_{n}\leq\letsymb{9}\}$ (consisting in the stochastic matrix of equation (\ref{dimension three circulant stochastic matrix})) in the first class do not satisfy the third characteristic property at $t=0$, regardless of the stochastic matrix. 
		\par \textit{Step 4.} Similarly, one can straightforwardly check through the respective matrix forms that the quantum operations $\{\Phi_{M}^{(r,v)},\;1\leq\letsymb{v}<r\leq\letsymb{4};\; 1\leq\letsymb{r}<v\leq\letsymb{4};\; 2\max(r,v)+1\leq\letsymb{M}_{n}\leq\letsymb{9} \}$ (consisting in the stochastic matrix of equation (\ref{dimension three circulant stochastic matrix}))   do not satisfy the third characteristic property at $t=0$, regardless of the stochastic matrix. 
		\par In addition, for $a=e=1/3,\;b=0,\;c=9/20,\;d=4/15,\; f=1/20$, while the stochastic matrix $\Lambda(t,0)$ is invertible for $0\leq\letsymb{t}<\infty$, the quantum operations $\{\Phi_{4}^{(1,1)},\Phi_{7}^{(3,3)},\Phi_{8}^{(2,2)},\Phi_{8}^{(3,3)}\}$, which are essentially the same,  do not also satisfy the third characteristic property for $t=\gamma^{-1}\ln(32+\frac{1+5\sqrt{673}}{4})$. Furthermore, for $a=c=e=0,\; b=0.25,\; d=0.1,\; f=0.2$, while the stochastic matrix  $\Lambda(t,0)$  is invertible for $0\leq\letsymb{t}<\infty$, the quantum operations $\{\{\Phi_{M}^{(1,1)}, 5\leq\letsymb{M}\leq\letsymb{9}\},\Phi_{5}^{(2,2)},\Phi_{7}^{(2,2)}, \Phi_{9}^{(2,2)},\Phi_{9}^{(4,4)}\}$, which are essentially the same, do not satisfy the third characteristic property for a time value $t\in(1.99393180,1.99393181)$, as one can numerically check  through the determinant of the respective matrix form. 
		\par Consequently, only the operations $\Phi_{3}^{(1,1)}$, $\Phi_{6}^{(2,2)}$ and $\Phi_{9}^{(3,3)}$  are remained to be checked. Due to Lemma \ref{lemma2 redundancy lemma},  $\Phi_{6}^{(2,2)}$ and $\Phi_{9}^{(3,3)}$ are simply the replications of $\Phi_{3}^{(1,1)}$.  Finally, according to Lemma \ref{lemma quantum operation inverse}, $\Phi_{3}^{(1,1)}$ is invertible, i.e. satisfies the third characteristic property, for $0\leq\letsymb{t}<\infty$ so long as  $\Lambda(t,0)$ is invertible.
	\end{proof}
	The Kraus operators in equation (\ref{Kraus operators}) might be linearly dependent for a particular stochastic matrix. For instance,  the identity matrix is a stochastic matrix at time $t=0$, and all of its corresponding Kraus operators are equal to itself. However, because of their structure, their linear dependence can only be of the form, $A_{j_{0}}=a_{1}A_{j_{1}}=\cdots=a_{k}A_{j_{k}}$ with $0\leq k \leq N-1$ such that all of the coefficients have norm one, $\abs{a_{i}}=1$. Since this is the case, the new set of linearly independent Kraus operators would be $\{\Tilde{A}_{j_{k}}, A_{j_{k+1}},\ldots, A_{j_{N-1}}\}$ with $\Tilde{A}_{j_{k}}=\sqrt{k+1}A_{j_{k}}$.
	\par Moreover, it could be argued that, given the non-uniqueness of the Kraus representation of a quantum operation, Theorems \ref{theorem1} and \ref{The first main theorem: uniqueness theorem} might be deemed inconclusive. However, this is not correct due to two reasons. First of all, we stress that according to Theorem \ref{The first main theorem: uniqueness theorem}, the Kraus representation of Theorem \ref{theorem1} is uniquely determined. Secondly, any other Kraus representation would be connected to that of Theorem \ref{theorem1} through a unitary matrix, $U=(u_{ms})_{N\times\letsymb{N}}$ \cite{Watrous2018}.  These two facts together imply that the Kraus operators of any Kraus representation of the quantum operation $\Phi_{c}(t,t_{1})$ have the form 
	\begin{equation}\label{Second Kraus operators form}
		B_{m}=\sum_{m=0}^{N-1}u_{ms}A_{s}, \quad m=0,1,\ldots, N-1,
	\end{equation}
	where $A_{s}$'s are the Kraus operators for $\Phi_{c}(t,t_{1})$ in equation (\ref{Kraus operators}). We also note that the matrix form of $\Phi_{c}(t,t_{1})$ is unique and determined by the Kraus operators of equation (\ref{Kraus operators}). Consequently, Theorems \ref{theorem1} and \ref{The first main theorem: uniqueness theorem}  are conclusive in the sense that the Kraus representation given in Theorem \ref{theorem1} is singled out uniquely. 
	\par Furthermore, one could argue that Theorems \ref{theorem1} and \ref{The first main theorem: uniqueness theorem} are inconclusive because of the other ways of representing quantum operation such as Natural and Choi representations. This objection would not be true because these representations are equivalent to the Kraus repsentation~\cite{Watrous2018}. 
	
	\section{Extension of classical P-divisibility}\label{sec:extension of P-divibility}
	In order to extend the classical P-divisibility to quantum processes, our point of departure is equation (\ref{stochastic to quantum representation}). To be more explicit, if a stochastic matrix $\Lambda(t,t_{1})$ is classical P-divisible, $\Lambda(t,t_{1})=\Lambda(t,s)\Lambda(s,t_{1})$, then the equation
	\begin{equation}\label{extension of P-divisibility first level}
		\Phi_{c}(t,t_{1})= \Phi_{c}(t,s)\circ \Phi_{c}(s,t_{1})
	\end{equation}
	must hold on the subset $\mathcal{D}(\mathcal{H}^{N})$, where $\Phi_{c}(t,t_{1})$, $\Phi_{c}(t,s)$ and $\Phi_{c}(s,t_{1})$ are the quantum operational representations of $\Lambda(t,t_{1})$, $\Lambda(t,s)$ and $\Lambda(s,t_{1})$, respectively, with the Kraus operators having the form given by equation (\ref{Kraus operators}). One can in fact check that equation (\ref{extension of P-divisibility first level}) holds on $\mathcal{D}(\mathcal{H}^{N})$. 
	\par Secondly, we show (see Theorem \ref{theorem4} below) that if the stochastic matrix $\Lambda(t,t_{1})$ is classical P-divisible,  its quantum operational representation $\Phi_{c}(t,t_{1})$ is CP-divisible, $\Phi_{c}(t,t_{1})=\Phi(t,s)\circ \Phi_{c}(s,t_{1})$, such that the action of $\Phi(t,s)$ on $\mathcal{D}(\mathcal{H}^{N})$ uniquely reduces to that of $\Phi_{c}(t,s)$. Therefore, we conclude that CP-divisibility for quantum operations is the natural extension of classical P-divisibility, and hence of the Chapman-Kolmogorov equation.
	
	\par Let us now consider that the stochastic matrix $\Lambda(t,t_{1})$ is invertible. Then,  Lemma \ref{lemma quantum operation inverse} and equation (\ref{principal submatrix decomposition of M}) yields   
	\begin{equation}\label{CP-divisible equation} \begin{split}
			M(t,s):&=M_{c}(t,t_{1})M_{c}^{-1}(s,t_{1})\\
			&=\Lambda(t,t_{1})\Lambda^{-1}(s,t_{1})\oplus V^{-1}(t,t_{1}) V(s,t_{1})  
		\end{split}
	\end{equation}
	for all  $t\geq s\geq t_{1}$ and $V(t,t_{1})=V_{1}(t,t_{1})\oplus\cdots\oplus\letsymb{V}_{N-1}(t,t_{1})$.  $M(t,s)$ can  be written in the form 
	\begin{equation}\label{third decomposition}
		M(t,s)=R(t,s)\odot G_{N}
	\end{equation} 
	where $R(t,s)$ consists of the matrices $\{V_{j}(t,t_{1})V_{j}^{-1}(s,t_{1}), j=0,1,\ldots,N-1\}$. For our purpose, we emphasize that $R(t,s)[\alpha_{0}]=\Lambda(t,s)$, where $\alpha_{0}$ is the index set introduced in Section \ref{sec:quantum operational representation}. We use this fact in Theorem \ref{theorem4}. 
	\begin{theorem}\label{theorem4}
		If the stochastic matrix $\Lambda(t,t_{1})=(\lambda_{ij})_{N\times\letsymb{N}}$ is (classical) P-divisible, then its quantum operational representation $\Phi_{c}(t,t_{1})$ is CP-divisible. 
	\end{theorem}
	\begin{proof}
		
		We give the proof for the invertible stochastic matrices. The result would also be valid for the noninvertible stochastic matrices, since every noninvertible matrix is the  limit of some invertible matrix \cite{HornJohnson2013}. Let us consider that $\Lambda(t,t_{1})$ is invertible and classical P-divisible for all $t\geq t_{1}$. Then, Lemma \ref{lemma quantum operation inverse} and equation (\ref{CP-divisible equation}) yield $M(t,s)=\Lambda(t,s)\oplus V(t,t_{1})V(s,t_{1})^{-1}$ with $\Lambda(t,s)$ being a stochastic matrix. Recalling that $M(t,s)$ is the matrix form of $\Phi(t,s)$, it is evident that the action of $\Phi(t,s)$ on the subset $\mathcal{D}(\mathcal{H}^{N})$ uniquely reduces to that of $\Phi_{c}(t,s)$.  
		\par According to Theorem $1$ of Ref. \cite{Poluikis1981},  quantum operations in $N$-dimensional Hilbert space are isomorphic to the cone of positive-semidefinite matrices $\mathcal{P}_{N}$.  As a result, it is sufficient for our purposes to show that $M(t,s)$ is isomorphic to a positive-semidefinite matrix. To this aim, we briefly introduce the notations used in Ref. \cite{Poluikis1981} to utilize the isomorphism. Let $\mathcal{K}_{n,m}$ ($\mathcal{K}_{n}$ if $m=n$) denote the vector space of $n\times\letsymb{m}$ matrices over the complex numbers.  Let $\mathcal{K}_{p,n}(\mathcal{K}_{q,m})$ be the collection of all $p\times\letsymb{n}$  block matrices with $q\times\letsymb{m}$ matrices as entries. $R=(r_{ij})\in \mathcal{K}_{pq,nm}$ may be written in the block form $R=(R_{ij})_{1\leq i\leq p;1\leq j\leq n}$, where $R_{ij}\in \mathcal{K}_{q,m}$ with $R_{ij}=(r^{ij}_{kl})_{q\times\letsymb{m}}$. We consider the set $S=\{(i,j)\mid i=0,1,\ldots,q-1;j=0,1,\ldots,n-1\}$ and the lexicographical ordering, $(i,j)<(r,s)$ iff $i<r$ or \mbox{($i=r$ and $j< s$)}, on $S$. The lexicographical ordering orders the elements of a matrix by rows, i.e. the first row entries first, the second row entries second, etc.  We also consider the bijection,  $[i,j]=in+j$, between $S$ and the set $\{0,1,\ldots, nq-1\}$ which corresponds to the lexicographical ordering. The bijective linear map $\Gamma:\mathcal{K}_{q^{2},n^{2}}\rightarrow\mathcal{K}_{q}(\mathcal{K}_{n})$ was introduced in Ref. \cite{Poluikis1981} as $\Gamma(R)^{ij}_{kl}=r_{[i,j][k,l]}$, and was shown later in Ref \cite{Barker1984} that $\Gamma$ is an isometrically isomorphism according to Hilbert-Schmidt inner product, i.e. for $\forall R,Q\in\mathcal{K}_{q^{2},n^{2}}$, $\langle \Gamma(R),\Gamma(Q)\rangle=\langle R,Q\rangle=\tr(R^{\dagger}Q)$.  $\Gamma$  isomorphically maps the matrix forms of quantum operations to positive semidefinite matrices \cite{Poluikis1981}.
		\par Since $M(t,s)$ is the matrix form of the  operation $\Phi_{c}(t,t_{1})\circ \Phi_{c}(s,t_{1})^{-1}$ and $M(t,s)\in \mathcal{K}_{N}(\mathcal{K}_{N})$, one can apply Theorem $1$ of Ref. \cite{Poluikis1981} to $M(t,s)$.  Hence, let $M(t,s)=R(t,s)\odot G_{N}$ as in equation (\ref{third decomposition}). We note that $\Gamma(G_{N})=G_{N}$ as one can directly check.  Since $\Gamma$ just reorders the elements of the matrices, we have  $\Gamma(M_{c}(t,s))=\Gamma(R(t,s)\odot G_{N})=\Gamma(R(t,s))\odot\Gamma(G_{N})=\Gamma(R(t,s))\odot G_{N}$. Recalling that $R(t,s)[\alpha_{0}]=\Lambda(t,s)$,  the diagonal elements of the matrix $\Gamma(R(t,s))\odot G_{N}$ become the elements of $\Lambda(t,s)$ such that the first $N$ diagonal elements are equal to the first row entries of $\Lambda(t,s)$, the second $N$ diagonal elements equal to the second row entries of $\Lambda (t,s)$, etc. Furthermore, $\Gamma(R(t,s))\odot\letsymb{G_{N}}$ is symmetric, implying that the eigenvalues are real. Since the summation of the eigenvalues of a matrix is equal to the summation of the diagonal elements, we conclude that the eigenvalues of  $\Gamma(R(t,s))\odot G_{N}$ are completely characterized by $\Lambda(t,s)$ so that their summation is nonnegative for any $\Lambda(t,s)$. This fact  implies that $\Gamma(R(t,s))\odot\Gamma(G_{N})$ is a positive semidefinite matrix concluding  that $M(t,s)$ must be a completely positive map. 
	\end{proof}
	In passing, we note that $\Phi(t,s)$ does not have to be equal to $\Phi_{c}(t,s)$. Nevertheless, if the stochastic matrix is circulant, then $\Phi(t,s)$  is equal to $\Phi_{c}(t,s)$ due to the fact that all of the principal submatrices $V_{j}$ in equation (\ref{principal submatrix decomposition of M})  become equal to each other. 
	\par Remarkably, Lemma \ref{lemma quantum operation inverse} and Theorem \ref{theorem4} establish the fact that CP-divisibility for quantum operations is the extension of classical P-divisibility, so thus of the Chapman-Kolmogorov equation. 
	\par Moreover, it is now evident that Theorems \ref{theorem1} and \ref{theorem4} together demonstrate that quantum states and quantum operations can be regarded as generalizations of one-point probability vectors and stochastic matrices, respectively. Indeed, the set of diagonal density matrices $\mathcal{D}(\mathcal{H}^{N})$, which is equivalent to the set of one-point probability vectors, is the subset of the set of quantum states $\mathcal{B}(\mathcal{H}^{N})$, and the set of quantum operational representations of the stochastic matrices $\{\Phi_{c}(t,t_{1})\}$ is the subset of the set of quantum operations $\{\Phi(t,t_{1})\}$ within finite-dimensional spaces. 
	\section{Application to dichotomic Markov Process}\label{sec: application}
	
	We illustrate our main results, namely Theorems \ref{theorem1} and \ref{theorem4}, by using two-state dichotomic Markov process which has been applied to many physical problems such as radiative transport problems \cite{Pomraning1986a, Pomraning1998a, Gao2020a} and random perturbation in the magnetic resonance \cite{Kubo1954a, Kampen2007, Balakrishnan2021}. Adopting the convention $\Lambda(t):=\Lambda(t,0)$, the stochastic matrix of the symmetric dichotomic Markov process with the transition rate $\gamma$ reads as \cite{Balakrishnan2021}   
	\begin{equation}
		\begin{split}
			\Lambda(t)=\frac{1}{2}\begin{pmatrix}
				1+\e^{-2\gamma t} & 1-\e^{-2\gamma t}\\
				1-\e^{-2\gamma t} & 1+\e^{-2\gamma t}
			\end{pmatrix}=\e^{-\gamma t}\begin{pmatrix}
				\cosh{\gamma t} & \sinh{\gamma t}\\
				\sinh{\gamma t} & \cosh{\gamma t}
			\end{pmatrix}.  
		\end{split}
	\end{equation}
	On using equation (\ref{Kraus operators}), the quantum operational representation $\Phi_{c}(t)$ has the Kraus operators 
	\begin{equation*}
	\ensuremath{\begin{split}
		A_{0}&=\frac{\e^{-\frac{\gamma t}{2}}}{\sqrt{2}}\begin{pmatrix}
				\sqrt{\cosh{\gamma t}} & \sqrt{\sinh{\gamma t}}\\
				\sqrt{\sinh{\gamma t}} & \sqrt{\cosh{\gamma t}}
			\end{pmatrix}, \\
			 A_{1}&=\frac{\e^{-\frac{\gamma t}{2}}}{\sqrt{2}}\begin{pmatrix}
				\sqrt{\cosh{\gamma t}} & -\sqrt{\sinh{\gamma t}}\\
				-\sqrt{\sinh{\gamma t}} & \sqrt{\cosh{\gamma t}}
			\end{pmatrix},  
		\end{split}}
	\end{equation*}
	which are linearly independent and satisfy the identity condition: $A^{\dagger}_{0}A_{0}+A^{\dagger}_{1}A_{1}=\mathit{I}_{2}$. According to equation (\ref{two dimensional representation matrix}) (or equation (\ref{matrix representation of quantum operation for stochastic matrices})), the matrix form of  $\Phi_{c}(t)$ is 
	\begin{equation}
		\begin{split}
			M_{c}(t)&=A_{0}\otimes\letsymb{\overline{A}}_{0}+A_{1}\otimes\letsymb{\overline{A}}_{1}\\
			&=\e^{-\gamma t}\begin{pmatrix}
				\cosh{\gamma t} & 0 & 0 & \sinh{\gamma t}\\
				0 & \cosh{\gamma t} & \sinh{\gamma t} & 0 \\
				0 & \sinh{\gamma t} & \cosh{\gamma t} & 0 \\
				\sinh{\gamma t} & 0 & 0 & \cosh{\gamma t}
			\end{pmatrix}
		\end{split}
	\end{equation}
	which admits the decomposition $M_{c}(t)=V_{0}(t)\oplus V_{1}(t)=\Lambda (t)\oplus \Lambda (t)$ in accordance with equation (\ref{principal submatrix decomposition of M}).
	\par The dichotomic Markov process is P-divisible, $\Lambda (t,s)=\Lambda(t)\Lambda(s)^{-1}$ for all $t\geq s\geq 0$, with 
	\begin{equation}
		\Lambda(t,s)=\e^{-\gamma (t-s)}\begin{pmatrix}
			\cosh{\gamma(t-s)} & \sinh{\gamma(t-s)}\\
			\sinh{\gamma(t-s)} & \cosh{\gamma(t-s)}
		\end{pmatrix},
	\end{equation}
	and its quantum operational representation is CP-divisible since 
	\begin{equation}
		\begin{split}
			M(t,s)&=M_{c}(t)M_{c}^{-1}(s)\!=\Lambda(t)^{-1}\Lambda(s)\!\oplus\Lambda(t)\Lambda^{-1}(s)\\
			&=\Lambda(t,s)\oplus \Lambda(t,s) 
		\end{split}  
	\end{equation}
	is isomorphic to a positive semidefinite matrix, as proved in Theorem \ref{theorem4}. We note that  $M(t,s)=M_{c}(t,s)$, and thus $\Phi(t,s)=\Phi_{c}(t,s)=\Phi_{c}(t)\circ \Phi_{c}(s)^{-1}$. The quantum operational representation of the dichotomic Markov process allows for the deformation of the process by incorporating certain additional quantum effects, which can be represented by additional Kraus operators. 
	\section{Conclusion}\label{sec:conclusion}
	We have shown that the relationship between classical stochastic processes and quantum processes cannot be considered merely a matter of some limits \cite{Kubo1992}; the physical state of a system does not only provide us with accessible information within the systems but also restricts the range of acceptable physical interactions which the system might experience. To state it more concretely, expressing the physical state of a classical stochastic process by a probability vector can give information about the system, thereby restricting the structure of dynamics represented by a stochastic matrix, i.e. the matrix must be nonnegative and its columns have to sum to unity. When the physical state of a system is represented by a quantum state $\rho$ instead of a probability vector $\mathbf{p}$, it allows us to consider a more general structure of dynamics given by a quantum operation.
	\par Theorem \ref{theorem1} embeds the time evolution of one-point probability distribution of the $N$-level stochastic processes into the quantum processes in $N$-dimensional Hilbert space.  Theorem \ref{The first main theorem: uniqueness theorem} establishes that the embedding by Theorem \ref{theorem1} is unique. Therefore, whenever the time evolution of the one-point probability distribution of stochastic processes is considered, quantum processes can be regarded as an extension of stochastic processes for finite dimensions. 
	\par Theorem \ref{theorem4} decisively demonstrates that CP-divisibility is the extension of classical P-divisibility, and therefore, a necessary condition for the quantum processes to be classified as Markovian. Consequently, the Markovianity criterion proposed by Breuer et al.~\cite{Breuer2009}, which is equivalent to quantum P-divisibility \cite{Weissman-Breuer-Vacchini2015}, does not align with the quantum extension of the Chapman–Kolmogorov equation. In contrast, the definition introduced by Rivas et al. \cite{Rivas2010} coincides precisely with CP-divisibility. However, recent studies \cite{McCauley2014, CanturkBreuer2024} have demonstrated the existence of classical P-divisible non-Markovian processes that are not mathematical artifacts but can be successfully applied to physical problems. For example, the discrete-time stochastic process constructed in Ref.~\cite{CanturkBreuer2024} is provably non-Markovian yet satisfies the Chapman–Kolmogorov equation (detailed in \ref{example for non-Markovian process}). Combining this with Theorems \ref{theorem1} and \ref{theorem4}, it follows that discrete-time CP-divisible non-Markovian quantum operations must also exist. Therefore, (at least for discrete-time evolution), CP-divisibility is a necessary but insufficient condition for quantum Markovianity—a conclusion also supported in earlier works \cite{Cruschinski2010, Benatti2012, Kumar2018}. We note that there are other proposals for the definition of the quantum Markovianity such as \cite{Alipour2012a, Luo2012, Utagi2021, Rajagopal2010a}, which deserve a separate analysis in regard to their relationships with CP-divisibility (for a recent review of the proposals, see Ref. \cite{Utagi2023a}).   
	\par Finally,  our work enables the employment of the quantum operations in stochastic processes. When this is the case, the physical state of the classical system of inquiry can be represented by a diagonal quantum state $\rho$ and the corresponding stochastic matrix with the Kraus operators in equation (\ref{Kraus operators}). This presents us with the possibility of generalizing the probability vector to a general quantum state and/or the quantum operational representation of the stochastic matrix to a relatively more general quantum operation.

 \begin{sloppypar}
	\Acknowledgement The author acknowledges financial support from TUBITAK (Project No: 124C348) and thanks G. Baris Bagci, Stephen R. Garcia, and Mehmet Küçükaslan for fruitful discussions and comments.   
\end{sloppypar}
 \begin{sloppypar}
\Conflictofinteres The author has no conflict of interest to declare that is relevant to the content of this article.
\end{sloppypar}
 \begin{sloppypar}
\DataAvailability Not applicable, as no new data were created or analyzed.
\end{sloppypar}

	\appendix 
	\section{The proof of Theorem \ref{theorem2}}\label{Proof_theorem_2}
	\begin{sloppypar}
	Let us consider a quantum operation $\Phi(t,t_{1})$ within $N$-dimensional Hilbert space with the corresponding Kraus operators, $\{R_{s}\}_{0\leq s\leq M-1}=\{(\Tilde{r}^{(s)}_{jk})_{N\times\letsymb{N}}\}_{0\leq s\leq M-1}$ and  the polar representation of the entries, $\Tilde{r}^{(s)}_{jk}:=(r^{(s)}_{jk})^{1/2}\e^{\mathrm{i}\phi_{0}(s,j,k)}$ so that $\mid\Tilde{r}^{(s)}_{jk}\mid^{2}=r^{(s)}_{jk}$. Applying the first condition of Definition \ref{definition3} to a general quantum state $\rho(t_{1})\in\mathcal{D}(\mathcal{H}^{N})$ yields
	\end{sloppypar}
	\begin{eqnarray*}
		\begin{split}
			&(\Phi(t,t_{1})\rho(t_{1}))_{jk}=\sum_{s=0}^{M-1}(R_{s}\rho(t_{1})R_{s}^{\dagger})_{jk}\\
			&\quad=\sum_{s=0}^{M-1}\sum_{l=0}^{N-1}\sqrt{r_{jl}^{(s)}r_{kl}^{(s)}}\rho_{ll}(t_{1})\e^{\mathrm{i}[\phi_{0}(s,j,l)-\phi_{0}(s,k,l)]}=\rho_{jj}(t)\delta_{jk},
		\end{split}
	\end{eqnarray*}
	which is possible only if
	\begin{enumerate}
		\item [P1.] $\displaystyle\sum_{s=0}^{M-1}\sqrt{r_{jl}^{(s)}r_{kl}^{(s)}}\e^{\mathrm{i}[\phi_{0}(s,j,l)-\phi_{0}(s,k,l)]}=\sqrt{\lambda_{jl}\lambda_{kl}}\delta_{jk}$,
	\end{enumerate}
	such that $\{\lambda_{km}\}$ form a stochastic matrix, $\Lambda(t,t_{1}):=(\lambda_{km})_{N\times\letsymb{N}}$. 
	\begin{enumerate}
		\item[P2.] For $j=k$, we obtain  $\sum_{s=0}^{M-1}r^{(s)}_{jl}=\lambda_{jl}$ from \mbox{P1}. 
	\end{enumerate}
	For $\forall j\neq k$, \mbox{P1} implies  
	\begin{enumerate}
		\item[P3.]   $\displaystyle\sum_{s=0}^{M-1}\sqrt{r_{jl}^{(s)}r_{kl}^{(s)}}\e^{\mathrm{i}[\phi_{0}(s,j,l)-\phi_{0}(s,k,l)]}=0$, $\forall j\neq k$, 
	\end{enumerate}
	which means that, taking into account \mbox{P2}, $\phi_{0}(s,j,l)-\phi_{0}(s,k,l)=\phi_{1}(s,j,k)$; in other words,
	\begin{enumerate}
		\item [P4.] $\phi_{0}(s,j,l)=f_{0}(s,j)+g_{0}(s,l)+\alpha$.
	\end{enumerate} 
	From the identity relation, $\sum_{s=0}^{M}R_{s}^{\dagger}R_{s}=I_{N}$, we obtain
	\begin{enumerate}
		\item[P5.] $\displaystyle\sum_{s=0}^{M-1}\sum_{l=0}^{N-1}\sqrt{r_{lj}^{(s)}r_{lk}^{(s)}}\e^{\mathrm{i}[\phi_{0}(s,l,k)-\phi_{0}(s,l,j)]}=\delta_{jk}$.
	\end{enumerate}
	For $j=k$, \mbox{P5} leads to  $\sum_{s=0}^{M-1}\sum_{l=0}^{N-1}r^{(s)}_{lj}=1$, which is consistent with \mbox{P2}. The right-hand side of \mbox{P5} is symmetric, so must be the left-hand side:
	\begin{enumerate}
		\item[P6.] $\sum_{s,l}\sqrt{r_{lj}^{(s)}r^{(s)}_{lk}}\e^{\mathrm{i}[g_{0}(s,k)-g_{0}(s,j)]}$$=\sum_{s,l}\sqrt{r_{lk}^{(s)}r^{(s)}_{lj}}\e^{-\mathrm{i}[g_{0}(s,k)-g_{0}(s,j)]}$,
	\end{enumerate}
	where we have used $\mbox{P4}$. We consider three cases separately: 
	\begin{description}
		\item[Case 1] Both $r_{jk}^{(s)}$ and phase function $\phi_{0}$ depend on $s$:
	\end{description}
	\mbox{P1}, \mbox{P5} and \mbox{P6} are possible only if $r^{(s)}_{jk}=\delta_{s,(Nj+k)}\lambda_{jk}$ with $s=0,1,2,\ldots,N^{2}-1$. In this case, $\phi(s,j,k)$ remains as an arbitrary global phase factor since the Kraus operators take on the form
	\begin{equation}\label{first class}
		R_{(N.j+k)}=\sqrt{\lambda_{jk}}\e^{\mathrm{i}\phi_{0}(Nj+k,j,k)}\ket{f_{j}}\bra{f_{k}}.
	\end{equation}
	\begin{description}
		\item[Case 2] $r_{jk}^{(s)}$ depends on $s$; phase function $\phi_{0}$ does not:
	\end{description}    
	This case is essentially the same as \textbf{Case 1} except that  $\phi_{0}$ is arbitrary in the form $\phi_{0}(j,k)$. More interestingly, below, we will show that a quantum operation demonstrated by the Kraus operators in equation (\ref{first class}) is essentially the same as one of the quantum operations in \textbf{Class 2}.
	
	\begin{description}
		\item[Case 3] Phase function $\phi_{0}$ depends on $s$; $r_{jk}^{(s)}$ does not:
	\end{description} 
	Then, $r^{(s)}_{jk}=r_{jk}$ and thus, \mbox{P2} reduces to 
	\begin{enumerate}
		\item [P2$^{\prime}$.] $\sum_{s=0}^{M-1}r_{il}=\lambda_{il}\Rightarrow r_{il}=\frac{\lambda_{il}}{M}$.
	\end{enumerate}
	Employing \mbox{P2$^{\prime}$} and \mbox{P4} in \mbox{P1} and \mbox{P5}, we obtain, respectively, 
	\begin{enumerate}
		\item [P7.]  $\frac{1}{M}\displaystyle\sum_{s=0}^{M-1}\e^{\mathrm{i}[f_{0}(s,i)-f_{0}(s,j)]}=\delta_{ij}, \;\;\frac{1}{M}\displaystyle\sum_{s=0}^{M-1}\e^{\mathrm{i}[g_{0}(s,i)-g_{0}(s,j)]}=\delta_{ij}$.    
	\end{enumerate}
	On using the identity in equation (\ref{number theory identity}), \mbox{P7} yields $\phi_{0}(s,j,k)= \alpha +2\pi s(r.j\pm v.k)/M$ for some fixed nonnegative integer numbers $r$ and $v$, and $N\leq M \leq N^{2}$. We omit the constant $\alpha$, since it does not have any physical significance. We can then rewrite the Kraus operators as follows:
	\begin{equation}\label{revised form of Kraus operators}
		R_{s}=(\Tilde{r}_{jk}^{(s)})_{N\times\letsymb{N}}, \;\Tilde{r}_{jk}^{(s)}= \frac{\sqrt{\lambda_{jk}}}{\sqrt{M}}\e^{\mathrm{i}(2\pi s(r.j\pm v.k)/M)}.
	\end{equation}
	The action of the Kraus operators in equation (\ref{revised form of Kraus operators}) on a general quantum state $\rho=(\rho_{jk})_{N\times\letsymb{N}}\in \mathcal{B}(\mathcal{H}^{N})$ yields
	\begin{equation}
	\ensuremath{\sum_{s=0}^{M-1}\left(R_{s}\rho R_{s}^{\dagger}\right)_{jk}=\sum_{s=0}^{M-1}\sum_{l,m=0}^{N-1}\frac{\sqrt{\lambda_{jl}\lambda_{km}}}{M}\rho_{lm}\e^{\frac{2\pi\mathrm{i}s}{M}(r(j-k)\pm v(l-m))}}.\label{application revised Kraus operators form to general quantum state} 
	\end{equation}
	On applying the first characteristic property in Definition \ref{definition3}, that is, for $\rho_{lm}=\rho_{ll}\delta_{lm}$, equation (\ref{application revised Kraus operators form to general quantum state}) reduces to 
	\begin{equation}
		\sum_{s=0}^{M-1}\left(R_{s}\rho R_{s}^{\dagger}\right)_{jk}=\sum_{s=0}^{M-1}\sum_{l=0}^{N-1}\frac{\sqrt{\lambda_{jl}\lambda_{km}}}{M}\rho_{ll}\e^{\frac{2\pi\mathrm{i}s}{M}r(j-k)}
	\end{equation}
	from which we obtain the condition $1\leq r(j-k)\mbox{mod}(M)< M$ to satisfy \mbox{P1}. Therefore, $r\geq 1$ and, without loss of generality, $r(j-k)< M\leq N^{2}$, which yields $r(N-1)<N^{2}\Rightarrow r<N+1+\frac{1}{N-1}$. Hence $1\leq r\leq N+1$. 
	\par In addition, the application of the second characteristic property in Definition \ref{definition3} to the quantum operation of the Kraus operators in equation (\ref{revised form of Kraus operators}) yields 
	\begin{equation*}
		\ensuremath{\sum_{s=0}^{M-1}\sum_{l,m=0}^{N-1}\sqrt{\lambda_{jl}\lambda_{jm}}\rho_{lm}\e^{\pm \frac{2\pi\mathrm{i}s}{M}v(l-m)} =\sum_{0}^{M-1}\sum_{l=0}^{N-1}\sqrt{\lambda_{jl}\lambda_{kl}}\rho_{ll}\e^{\frac{2\pi\mathrm{i}s}{M}r(j-k)},}
	\end{equation*}
	which is valid only if $1\leq v(l-m)\mbox{mod}(M)<M$. Similar to the case for $r$, we obtain $1\leq v \leq N+1$.  
	\par On the other hand, since $r(N-1)< M$ and $v(N-1)<M$, the condition $\max(r,v)(N-1)+1\leq M \leq N^{2}$ must hold. 
	\par To conclude, we are left with two classes of quantum operations in an $N$-dimensional Hilbert space. The first class is \textbf{Class 1} having the following sets of quantum operations
	\begin{equation}
		C_{1}^{rv}=\{\Psi_{M_{1}}^{rv},\Psi_{M_{2}}^{rv},\ldots,\Psi_{N^{2}}^{rv}\};\; 1\leq r,v\leq N+1 
	\end{equation}
	such that for fixed values of $r$ and $v$ the quantum operation $\Psi^{rv}_{M_{n}}$ is determined by the following set of Kraus operators
	\begin{equation}
		T_{s}=(t_{jk})_{N\times\letsymb{N}}, \; t_{jk}=\frac{\sqrt{\lambda_{jk}}}{\sqrt{M_{n}}}\e^{\frac{2\pi\mathrm{i}s}{M_{n}}(rj+vk)},
	\end{equation}
	where  $s=0,1,\ldots, M_{n}-1$, and  $M_{n}=\max (r,v)(N-1)+n$ with $1\leq n \leq N^{2}-\max(r,v)(N-1)$. Similarly, the second class is \textbf{Class 2} consisting of the following sets of quantum operations
	\begin{equation}
		C_{2}^{rv}=\{\Phi_{M_{1}}^{rv},\Phi_{M_{2}}^{rv},\ldots,\Phi_{N^{2}}^{rv}\};\; 1\leq r,v\leq N+1 
	\end{equation}
	such that for fixed values of $r$ and $v$ the quantum operation $\Phi_{M_{n}}^{rv}$ is determined by the following set of Kraus operators 
	\begin{equation}
		R_{s}=(r_{jk})_{N\times\letsymb{N}}, \; r_{jk}=\frac{\sqrt{\lambda_{jk}}}{\sqrt{M_{n}}}\e^{\frac{2\pi\mathrm{i}s}{M_{n}}(rj-vk)},
	\end{equation}
	where  $s=0,1,\ldots, M_{n}-1$, and  $M_{n}=\max (r,v)(N-1)+n$ with $1\leq n \leq N^{2}-\max(r,v)(N-1)$. 
	\par Now, we show that the quantum operation $\mathcal{E}$ defined by the Kraus operators in equation (\ref{first class}) is essentially the same as $\Phi_{N^{2}}^{(N,1)}$. Without loss of generality, let us rewrite the Kraus operators in equation (\ref{first class}) as $K_{N.j+k}=\sqrt{\lambda_{jk}}\ket{e_{j}}\bra{e_{k}}$. Note that the Kraus operators of the quantum operation $\Phi_{N^{2}}^{(N,1)}$ take on the form 
	\[R_{s}=(r_{jk})_{N\times\letsymb{N}}, \; r_{jk}=\frac{\sqrt{\lambda_{jk}}}{N}\e^{\frac{2\pi\mathrm{i}s}{N^{2}}(Nj-k)}.\]
	Then, defining the unitary matrix \[U=(u_{sp})_{N^{2}\times\letsymb{N^{2}} }, \; u_{s(N.j+k)}:=\frac{1}{N}\e^{\frac{2\pi\mathrm{i}s(N.j-k)}{N^{2}}}\]
	with $s=0,1,\ldots, N^{2}-1$ and $j,k=0,1,\ldots,N-1$, it is  evident that $R_{s}=\sum_{j,k=0}^{N-1}u_{s(N.j+k)}K_{N.j+k}$.
	Therefore, $\mathcal{E}$ and $\Phi_{N^{2}}^{(N,1)}$ are essentially the same since $U$ is an isometry. Defining $V=(v_{jk})_{N^{2}\times\letsymb{ N^{2}}}:=U^{\dagger}$, one can equivalently write $K_{N.l+m}=\sum_{j,k=0}^{N-1}v_{(N.l+m)(N.j+k)}R_{N.j+k}$.\qed
	\section{P-divisible Non-Markovian Process}\label{example for non-Markovian process}
	 While P-divisible non-Markovian processes exist (see Ref.~\cite{CanturkBreuer2024} and references therein), we briefly review the explicit example constructed by Canturk and Breuer~\cite{CanturkBreuer2024}. Consider a discrete-time stochastic process $X_{t}$ with sample space $\mathbb{E}=\{0,1\}$. For the ordered time instants $t_{1}<t_{2}<\ldots<t_{3k}$ (with $k=1,3,4,\ldots,n$), the authors assume that joint probability of order $3k$  factorizes as 
	\begin{equation*}\label{joint probability of 3k hierarchy}
		\begin{split}
			&p_{3k}(j_{3k},t_{3k};j_{3k-1},t_{3k-1};j_{3k-2},t_{3k-2};\ldots;j_{3},t_{3};j_{2},t_{2};j_{1},t_{1})\\
			&=p_{3}(j_{3k},t_{3k};j_{3k-1},t_{3k-1};j_{3k-2},t_{3k-2})\ldots p_{3}(j_{3},t_{3};j_{2},t_{2};j_{1},t_{1}),      
		\end{split}
	\end{equation*}
	 such that \[p_{3}(j_{3k},t_{3k};j_{3k-1},t_{3k-1};j_{3k-2},t_{3k-2})=p_{3}(j_{3},t_{3};j_{2},t_{2};j_{1},t_{1}).\] Then, for the discrete time instants $t_{1}<\letsymb{t_{2}}<\letsymb{t_{3}}$ and the initial probability $\boldsymbol{q}(t_{1})=(q_{1},q_{2})$, the authors construct the following joint probability of order three
	\begin{equation*}\label{Two state non-Markovian process}
		\begin{split}
			p_{3}&(j_{3},t_{3};j_{2},t_{2};j_{1},t_{1})=\frac{\varepsilon\letsymb{q_{1}}}{2}\delta_{j_{3},0}\delta_{j_{2},0}\delta_{j_{1},0}+\frac{\varepsilon}{2}q_{1}\delta_{j_{3},1}\delta_{j_{2},1}\delta_{j_{1},0}\\
			&+\frac{1-\varepsilon}{2}q_{1}\delta_{j_{3},1}\delta_{j_{2},0}\delta_{j_{1},0}+\frac{1-\varepsilon}{2}q_{1}\delta_{j_{3},0}\delta_{j_{2},1}\delta_{j_{1},0}\\
			&+\frac{q_{2}}{2}\delta_{j_{3},0}\delta_{j_{2},1}\delta_{j_{1},1}+\frac{q_{2}}{2}\delta_{j_{3},1}\delta_{j_{2},0}\delta_{j_{1},1} , \quad j_{3}, j_{2},j_{1}\in\mathbb{E},
		\end{split}
	\end{equation*}
where $\delta_{j,k}$ is the Kronecker delta and  $\varepsilon \in\left [0,1\right]$ is memory parameter. The process reduces to a Markov process for $\varepsilon=0$. Despite its non-Markovian character for $\varepsilon>0$, it is shown that this process still satisfies the Chapman–Kolmogorov equation.

	%
	
	\end{document}